\documentclass[journal]{IEEEtran}

\usepackage{cite}
\usepackage[cmex10]{amsmath}
\usepackage{textcomp}
\usepackage{hyperref}
\usepackage[utf8]{inputenc}
\usepackage{graphicx}
\graphicspath{{Figures/}}
\usepackage{epsfig}
\usepackage{color}

\definecolor{blue}{rgb}{0,0,1}
\definecolor{green}{rgb}{0,0.5,0}
\definecolor{red}{rgb}{1,0,0}
\definecolor{pink}{rgb}{0.9,0.3,0.7}
\definecolor{azur}{rgb}{0,0.5,0.5}
\definecolor{orange}{rgb}{1,0.5,0.2}
\definecolor{brown}{rgb}{0.5,0,0}

\newcommand{\be}{\begin{equation}}
\newcommand{\ee}{\end{equation}}
\newcommand{\ben}{\begin{equation*}}
\newcommand{\een}{\end{equation*}}
\newcommand{\ba}{\begin{eqnarray}}
\newcommand{\ea}{\end{eqnarray}}
\newcommand{\ban}{\begin{eqnarray*}}
\newcommand{\ean}{\end{eqnarray*}}

\newcommand{\ie}{\textit{i.e.} }

\newcommand{\leg}[1]{\textbf{#1}}

\begin{document}

\markboth{IEEE Transactions on Robotics}{Special Issue on Robotic Sense of Touch}

\title{Perception of Surface Defects by Active Exploration with a Biomimetic Tactile Sensor}

\author{Raphaël~Candelier,~Georges~Debrégeas~and~Alexis~Prevost
  \thanks{ Laboratory of Statistical Physics, CNRS-ENS-UPMC UMR8550, Ecole Normale Supérieure, 24 rue Lhomond 75005 Paris, France. e-mail: (see http://www.lps.ens.fr/recherche/systemes-biologiques-integres/).}%
  \thanks{Manuscript received ...; revised ...}
}

\maketitle

\begin{abstract}
We investigate the transduction of tactile information during active exploration of finely textured surfaces using a novel tactile sensor mimicking the human fingertip. The sensor has been designed by integrating a linear array of 10 micro-force sensors in an elastomer layer.  We measure the sensors' response to the passage of elementary topographical features in the form of a small hole on a flat substrate. The response is found to strongly depend on the relative location of the sensor with respect to the substrate/skin contact zone. This result can be quantitatively interpreted within the scope of a linear model of mechanical transduction, taking into account both the intrinsic response of individual sensors and the context-dependent interfacial stress field within the contact zone. Consequences on robotics of touch are briefly discussed.
\end{abstract}


\section{Introduction}


\IEEEPARstart{T}{he} tactile sensitivity of hands and fingertips allows humans to dexterously manipulate objects and obtain a wealth of information such as their shape, weight or texture~\cite{hollins2009som}. Specialized nerve endings (mechanoreceptors) located in the first layers of the skin trigger sequences of spikes traveling up the afferent fibers, that are further processed by the central nervous system to ultimately form a representation of the probed object. Biomimetic tactile sensing aims at artificially reproduce this mechanism by mimicking the fingerpad. Biomimetic sensors generally consist of a collection of force sensors covered with an elastic membrane ~\cite{fearing1990tsm, hosoda2006ars,dahiya2010tsf}. Different technologies have been used to produce the artificial counterparts of the mechanoreceptors, such as MEMS\cite{scheibert2009rof, sunghoon2005tcu, wisitsoraat2006daf, muhammad2009dob}, strain-sensitive materials~\cite{howe1993dts, derossi2005pbi, lacour2004aes} or thin-film devices~\cite{maheshwari2006hrt}. More recently, the role of the skin itself in the transduction process has been studied. It has been shown in particular that the topography of the skin may significantly modify the sensors response ~\cite{derossi2005pbi, scheibert2009rof}.

Despite the rapid development of this field in recent years,
replication of the full manipulative capabilities of the human hand is still years away. Beyond the technological challenge of designing realistic tactile devices, efforts are still needed to understand how one can efficiently and robustly extract physical information about the probed object from the sensors signal. One of the fundamental difficulties owes to the fact that tactile perception is an active process: the sensor's response not only depends on the intrinsic property of the sensor system but also on the exploratory conditions, \ie the way the object is probed. This may explain why, in natural tactile perception, various types of exploratory procedures are spontaneously used depending on the type of information one seeks to extract (shape, texture, weight, etc.)~\cite{Lederman1974tro}. How these various exploratory conditions shape the elicited mechanoreceptors response and thus effectively participate to the transduction and encoding of tactile information is to date mostly unknown.

In this Letter we focus on the role of one exploratory parameter, namely the location of the fingertip/surface contact zone with respect to the sub-cutaneous receptors. To that end, we investigate the response of a multi-sensor biomimetic device to the passage of elementary topological defects and examine how this response varies with the location of each sensor below the contact zone.

\section{Biomimetic fingertip design and calibration}

\begin{figure}[!b]
\center
\includegraphics[width=0.70\columnwidth]{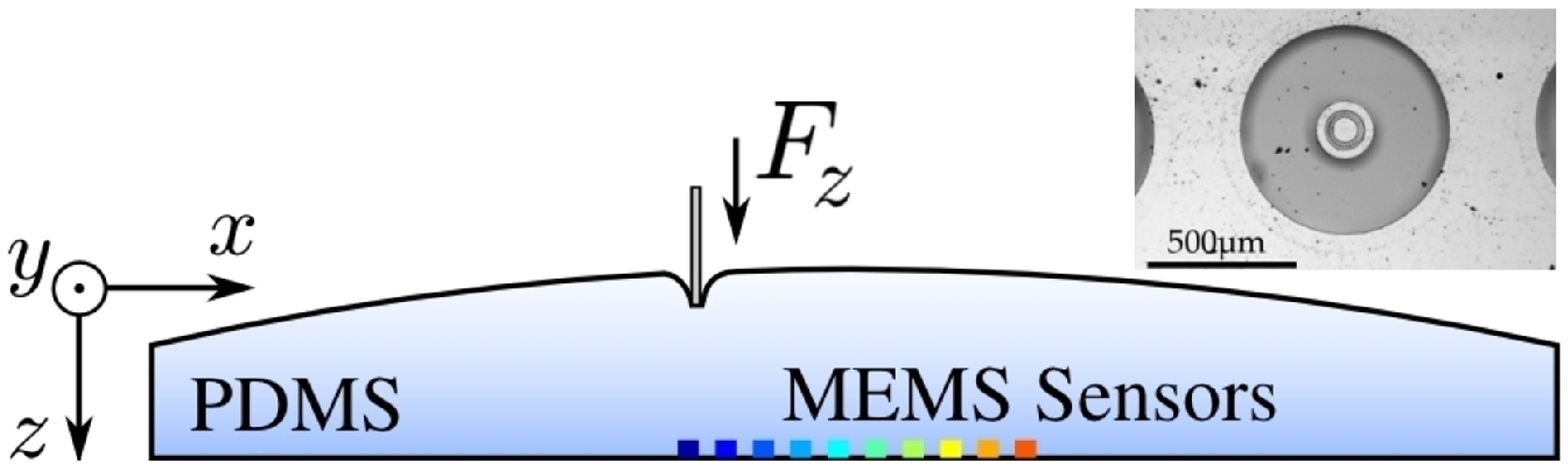}
\includegraphics[height=0.48\columnwidth]{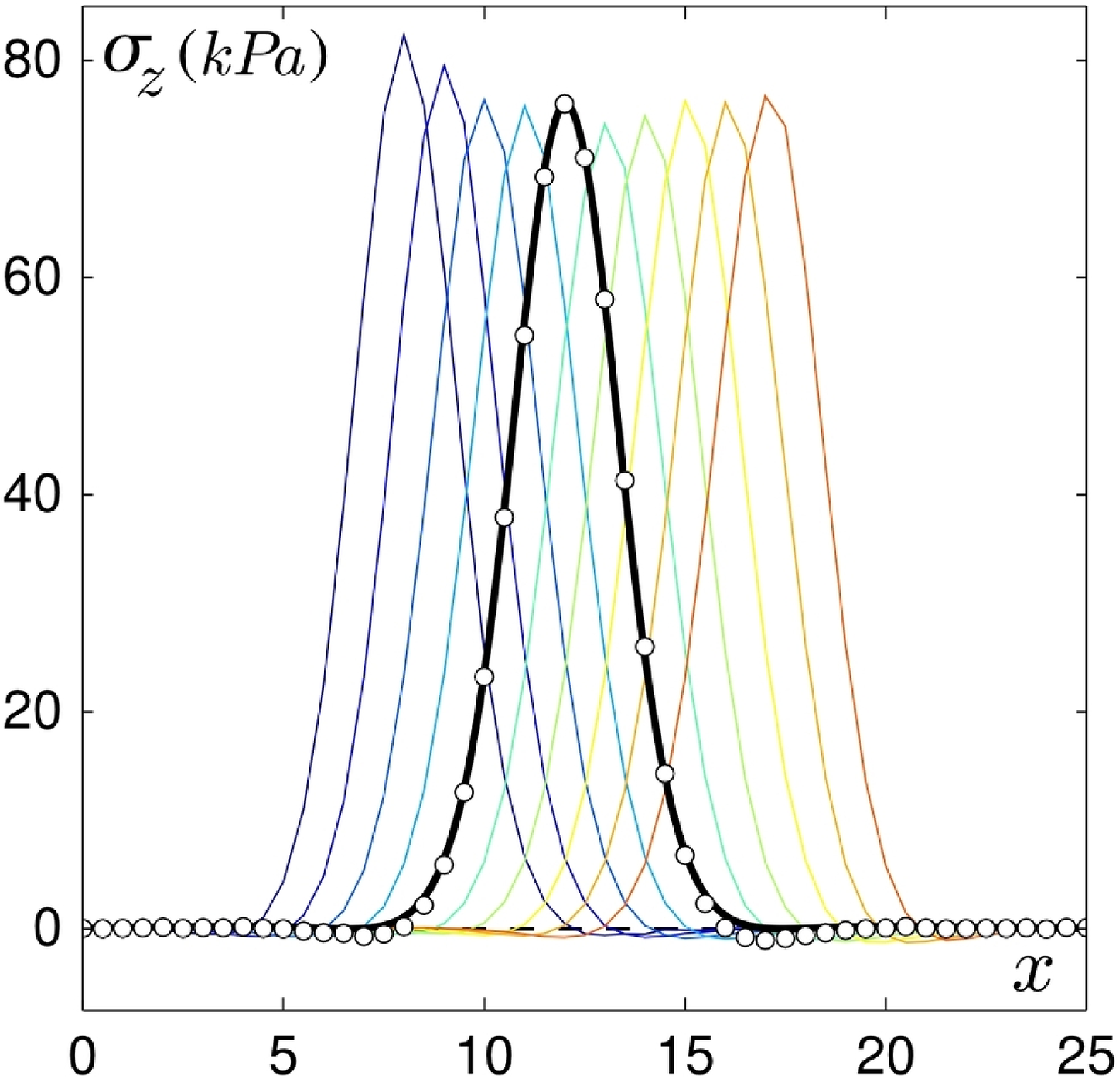}
\includegraphics[height=0.48\columnwidth]{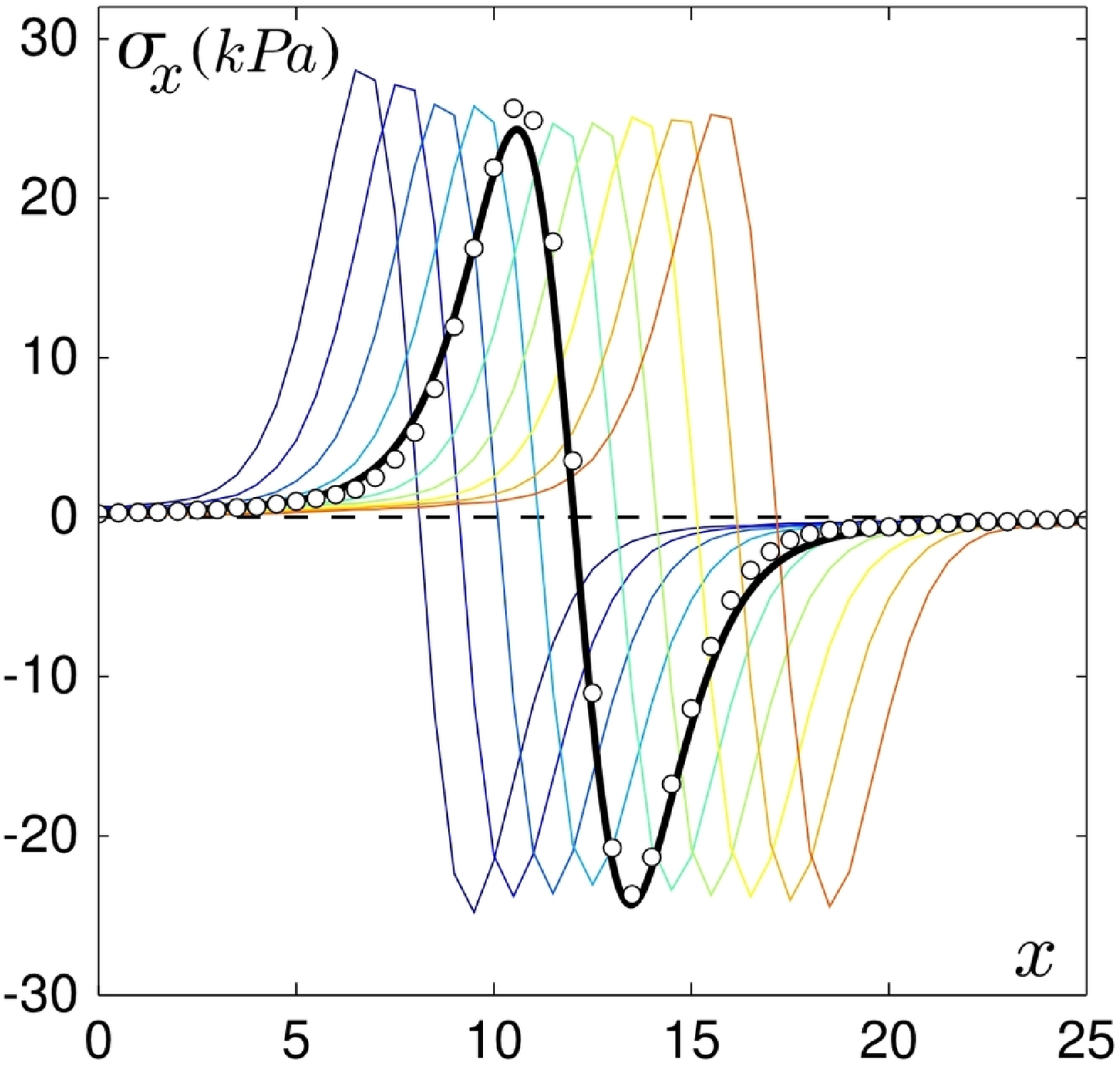}
\caption{ Calibration of the biomimetic sensors.
\leg{Top} Sketch of the experimental system along with the indentation protocol: a line of $10$ MEMS sensors are embedded in a spherical elastomer cap. It allows for the measurement of the subcutaneous stress induced by a local indentation with a wire of diameter $500\mu m$. \textit{Inset} Image as seen from above of one of the 10 MEMS sensors. 
\leg{Bottom} Normal (right) and tangential (left) stress $\sigma_z$ and $\sigma_x$ measured by the $10$ micro-force sensors as a function of the indentor's position $x$ ($y=0$) for a normal force $F_z=1N$. The stress measured by sensor $\sharp5$ (circles) is compared to the prediction of a finite element calculation (thick black curve).
}
\label{fig:system_cal}
\end{figure}

A novel biomimetic tactile sensor has been designed with the aim to reproduce the functioning principle and main mechanical characteristics of the human fingertip. The force sensing device is a linear array of 10 MEMS force sensors\footnote{The MEMS sensors have been developed by Patrice Rey and collaborators at CEA-LETI in Grenoble, France.}. Successive sensors are $1mm$ apart and aligned along the $x$-axis (see fig.\ref{fig:system_cal}-top), which is the scanning direction in dynamic sensing experiments.  Sensors are numbered from left to right and a consistent color code is used in all figures, from blue ($1$) to red ($10$). The sensitive part of each force sensor consists of a vertical silicon cylinder of diameter $200\mu m$ and height $400\mu m $ attached to an horizontal suspended silicon membrane of radius $350\mu m$ (see inset in fig.\ref{fig:system_cal}-top). This membrane bears 4 pairs of piezoresistive gauges that measure its internal stress state, from which one can extract the three components of the force acting on the cylinder.

The array of force sensors is embedded in an elastic spherical cap mimicking the skin, with a radius of curvature $R=129.7 mm$ and a maximum thickness $h=3.04mm$. The apex of the elastic layer is positioned roughly above the middle of the array. This elastic cover is made of cross-linked PolyDiMethylSiloxane (PDMS, Sylgard 184, Dow Corning) whose Young's modulus and Poisson ratio are respectively $\sim$ $3 MPa$ and $0.5$. Its surface is rendered rough at a micrometric scale by mechanically abrading the spherical mold used for its design. This fine texturing significantly reduces adhesive and frictional forces which may otherwise result in the deterioration of the sensor. 

The biomimetic sensor is integrated in a mechanical set-up which allows for controlled tactile stimulation. The probed substrate is attached to a three-axis motorized platform through a double cantilever system coupled with capacitive position sensors that enable one to simultaneous record the global normal force $F_z$ and tangential force $F_x$.

Individual micro-force sensors are first calibrated using a standard quasi-punctual indentation protocol~\cite{howe1993dts, scheibert2008EPL} sketched in fig.\ref{fig:system_cal}-top. A $500 \mu m$ in diameter metallic wire is indented normally to the skin at various positions under a controlled load $F_z$ in the range $0-0.4N$. Regardless of the indentor's location, the sensors' response varies linearly with the applied load. Figure~\ref{fig:system_cal}-bottom displays this response for the ten sensors corresponding to a unit load ($F_z=1N$) as a function of the position $x$ of the indentor, for both the normal $\sigma_z$ and tangential $\sigma_x$ components.

These point-load response profiles are compared with the result of a finite element simulation using CAST3M\footnote{CAST3M is an open source finite element simulation software. See \url{http://www-cast3m.cea.fr/cast3m/index.jsp}}: the stress field at the bottom of an elastic layer of finite thickness $h$ rigidly attached to a solid substrate is calculated as its surface is normally indented by a $500\mu m$ flat punch. In the limit where the punch diameter is much smaller than the layer thickness, the calculated stress field is similar to that obtained for a punctual indentor (the so-called Green function) and is a function of $h$ only. As shown in the bottom panels of figure~\ref{fig:system_cal}, the measured sensors' response profiles and calculated stress fields can be correctly adjusted for both directions of the stress. This adjustment provides a Volt-to-Pascal calibration of the sensors, as well as an estimated value of the thickness of the layer at each sensor's location. The latter is consistent with the known geometry of the elastic cap: in particular, the layer thickness is found to vary continuously along the line of sensors as expected for a spherical cap, with a maximal thickness of the skin located close to sensor $\sharp6$.  
This thickness variation can be clearly seen on fig.~\ref{fig:system_cal} through an inverse variation of the maxima of $\sigma_z$ (corresponding to the point where the indentor lays right above the sensor).

\section{Stress field during exploration of a smooth substrate}

\begin{figure}[!t]
\center
\includegraphics[width=0.70\columnwidth]{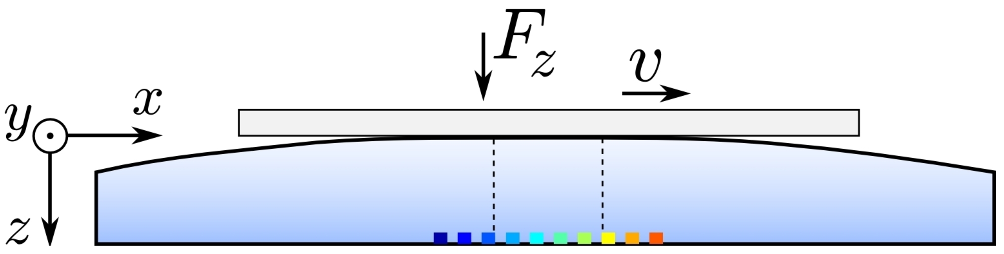}
\includegraphics[height=0.48\columnwidth]{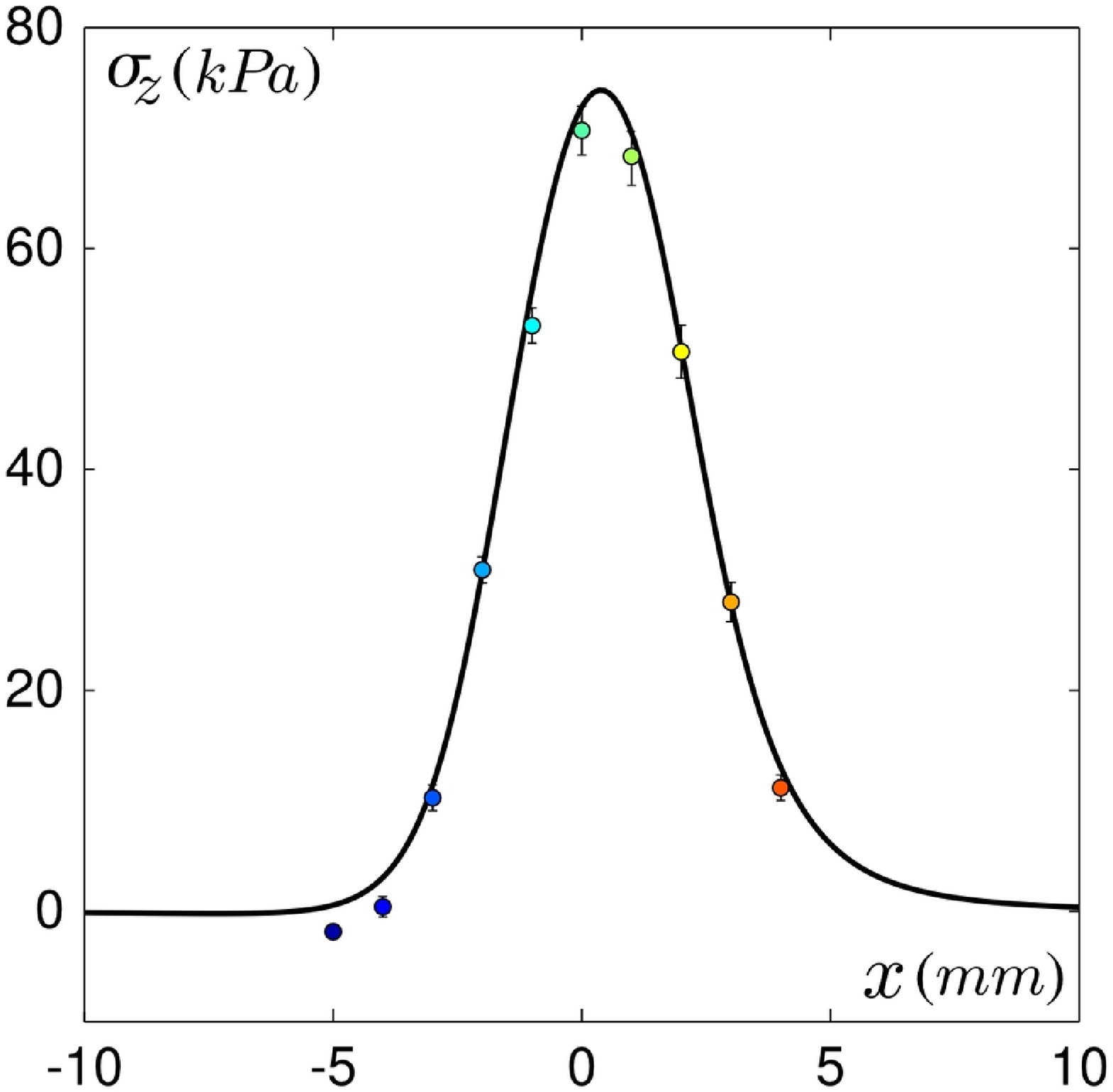}
\includegraphics[height=0.48\columnwidth]{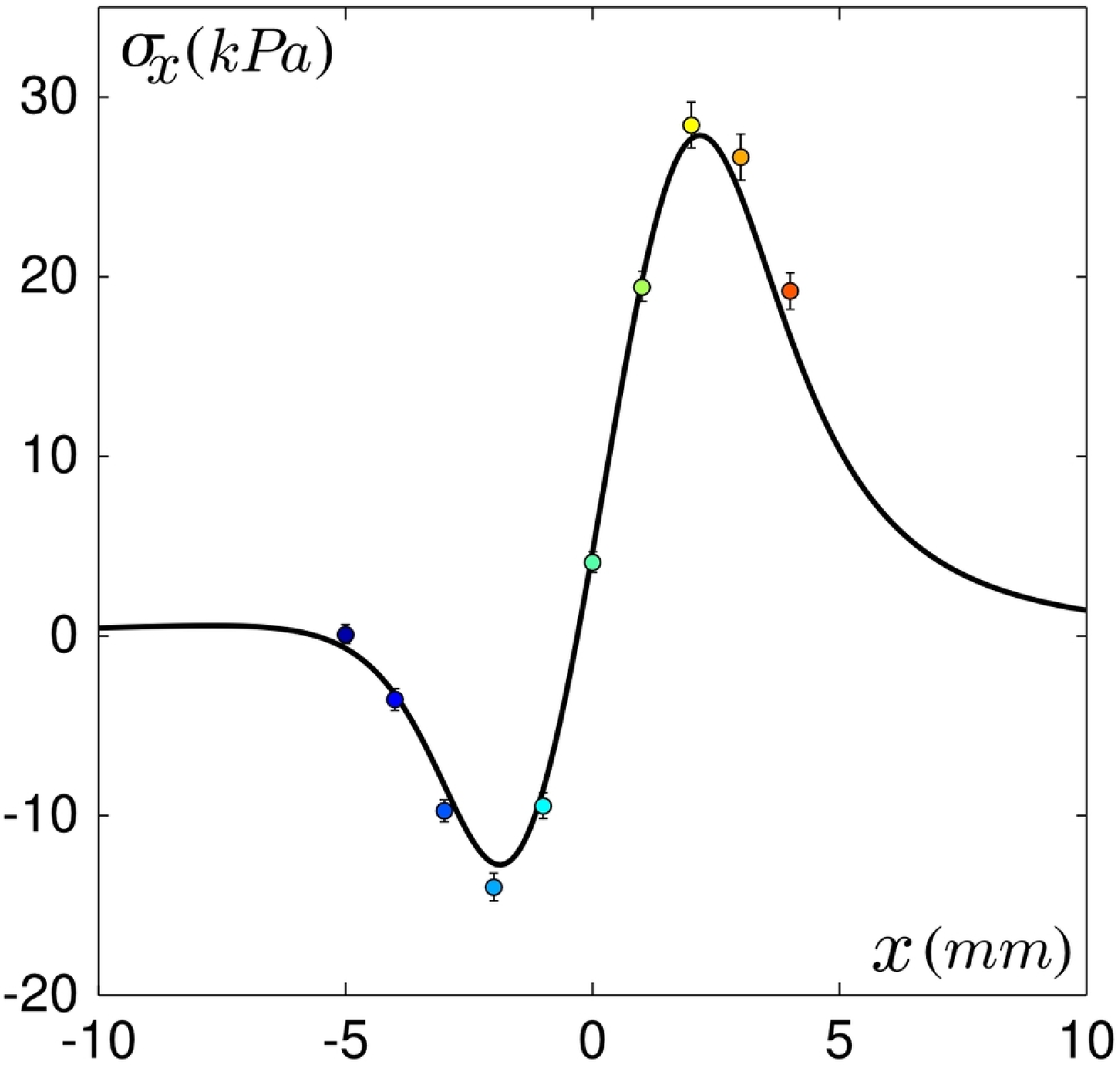}
\includegraphics[height=0.48\columnwidth]{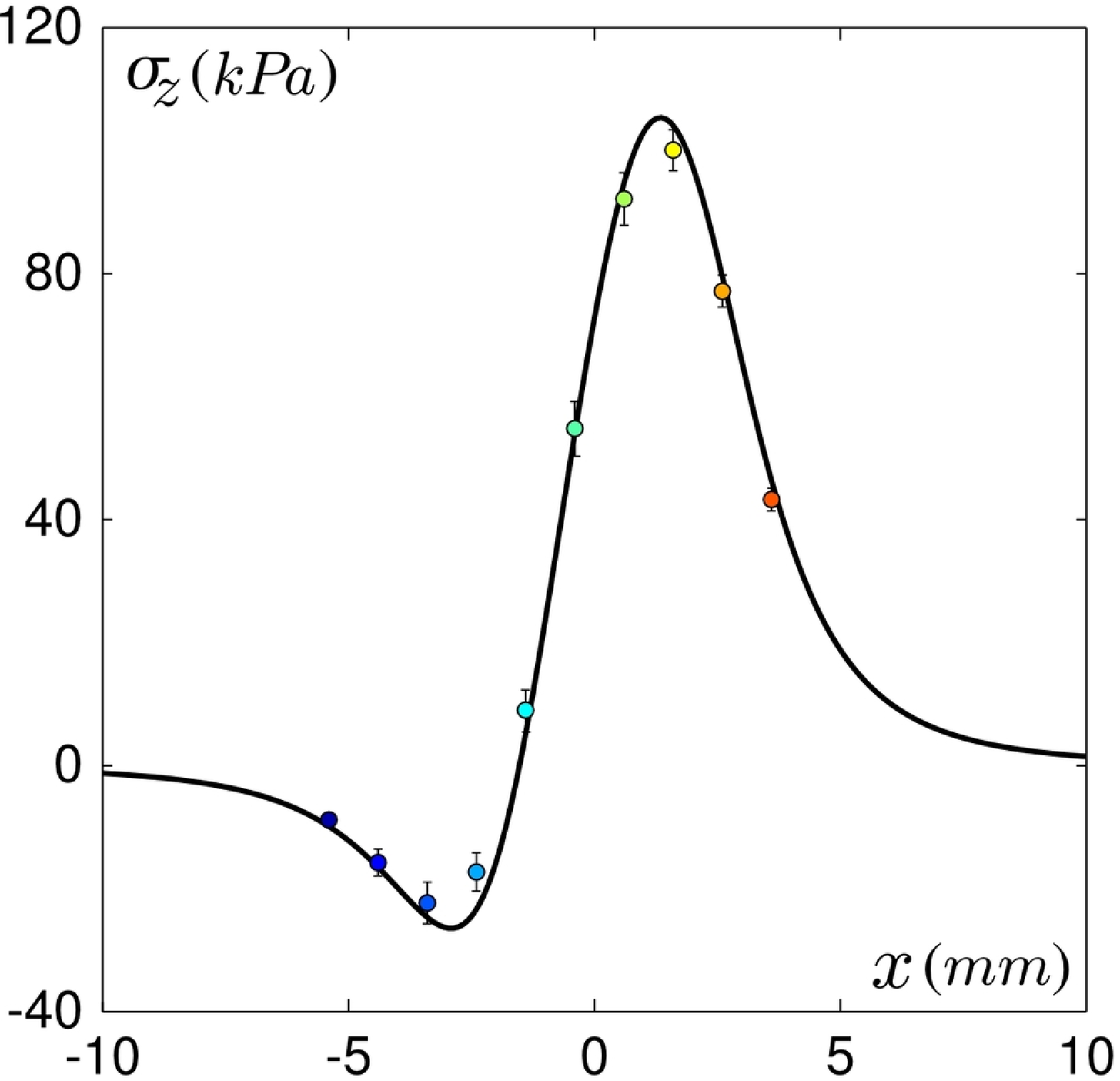}
\includegraphics[height=0.48\columnwidth]{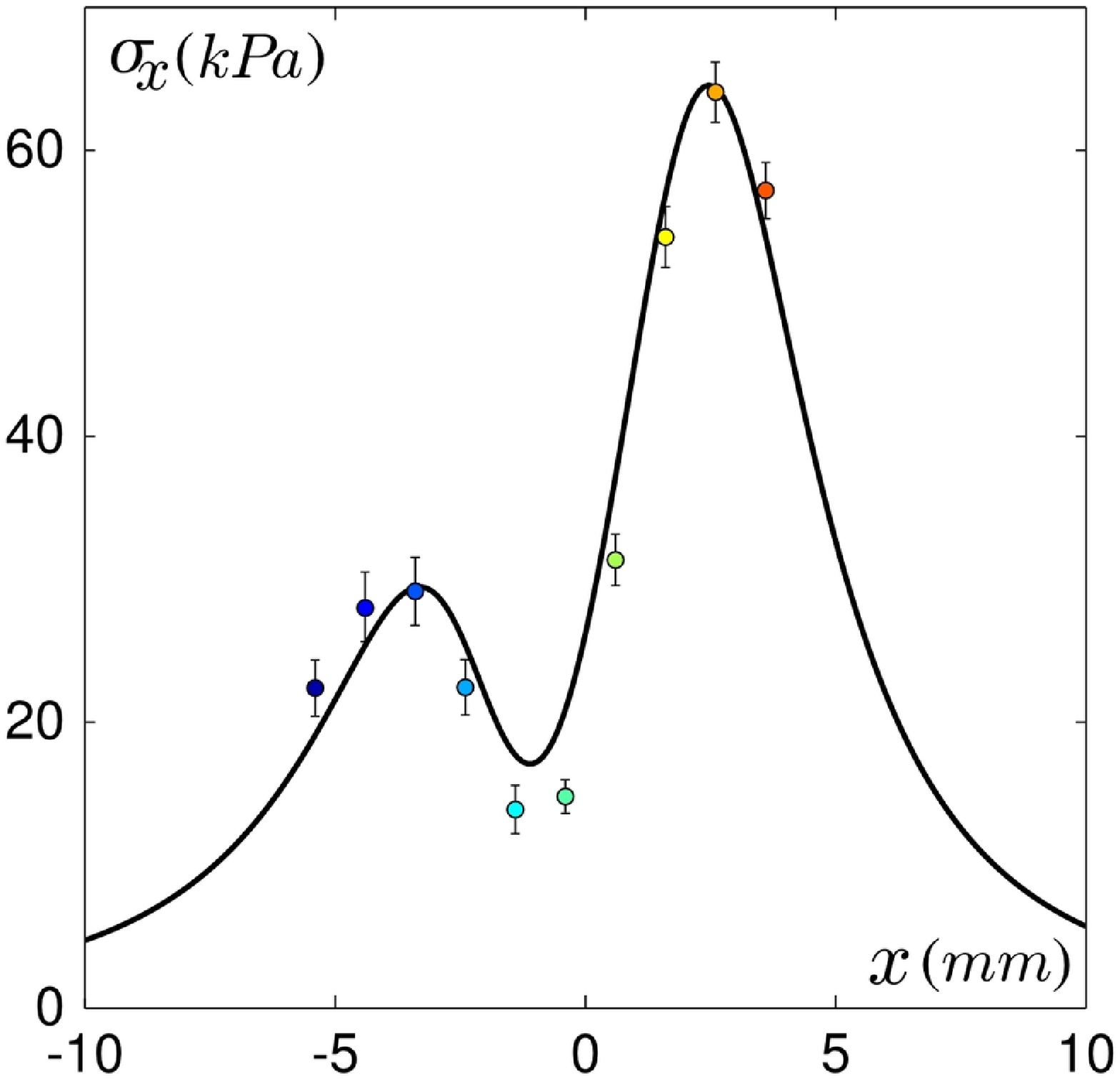}
\caption{ Static and dynamic average stress as measured by the sensors compared to the model prediction.
\leg{Top} Sketch of the experiment: a smooth substrate is pushed against the skin surface with a confining normal load $F_z = 0.8N$. In dynamic exploration, the substrate moves with a velocity $v$ along the sensor array axis.
\leg{Middle} Measured average stress in the static regime for the 10 sensors (circles) compared to the model's prediction (black curve) for the normal component $\sigma_z$ (left) and the tangential component along the direction of motion $\sigma_x$ (right), as functions of the sensor's location $x$. Position $x=0$ is the center of the contact. Error bars show the measured standard deviation and colors correspond to the locations depicted on the sketch.
\leg{Bottom} Same as above but in the steady sliding regime ($v=500\mu m.s^{-1}$).
}
\label{fig:avg_stress}
\end{figure}

We first investigate the response of the biomimetic sensor to a flat and smooth substrate, under both static and dynamic exploration. A Plexiglas plate is put into contact with the sensor under a constant load ($F_z=0.8N$) forming a circular contact of diameter $4-5mm$. The interfacial contact stress field is in this case centro-symmetric with respect to the center of the contact zone. The corresponding normal and tangential stress $\sigma_z$ and $\sigma_x$,  measured with the micro-force sensors, display respectively symmetric and anti-symmetric profiles around the contact center, as shown in figure~\ref{fig:avg_stress}-middle. A slight asymmetry in the $x$-direction is however observed which can be accounted for by the development of a minute tangential force under normal loading associated with the presence of the cantilever~\cite{scheibert2008EPL}. The substrate is then put into sliding motion along the $x$-axis at constant velocity $v=500 \mu m.s^{-1}$. The time-averaged profiles in steady sliding are shown in the bottom panel of figure~\ref{fig:avg_stress}. These profiles strongly differ from their static counterpart as a result of the development of friction-induced tangential stress in the contact zone.

In order to interpret the sensors' response in both static and dynamic conditions, a simple mechanical model of linear elastic transduction is developed. Three ingredients are needed: the pressure field at the skin/substrate interface, a local law of friction to derive the tangential stress in the contact zone and a description of the stress propagation in an elastic layer of finite thickness.

\begin{figure}[!t]
\center
\includegraphics[height=0.48\columnwidth]{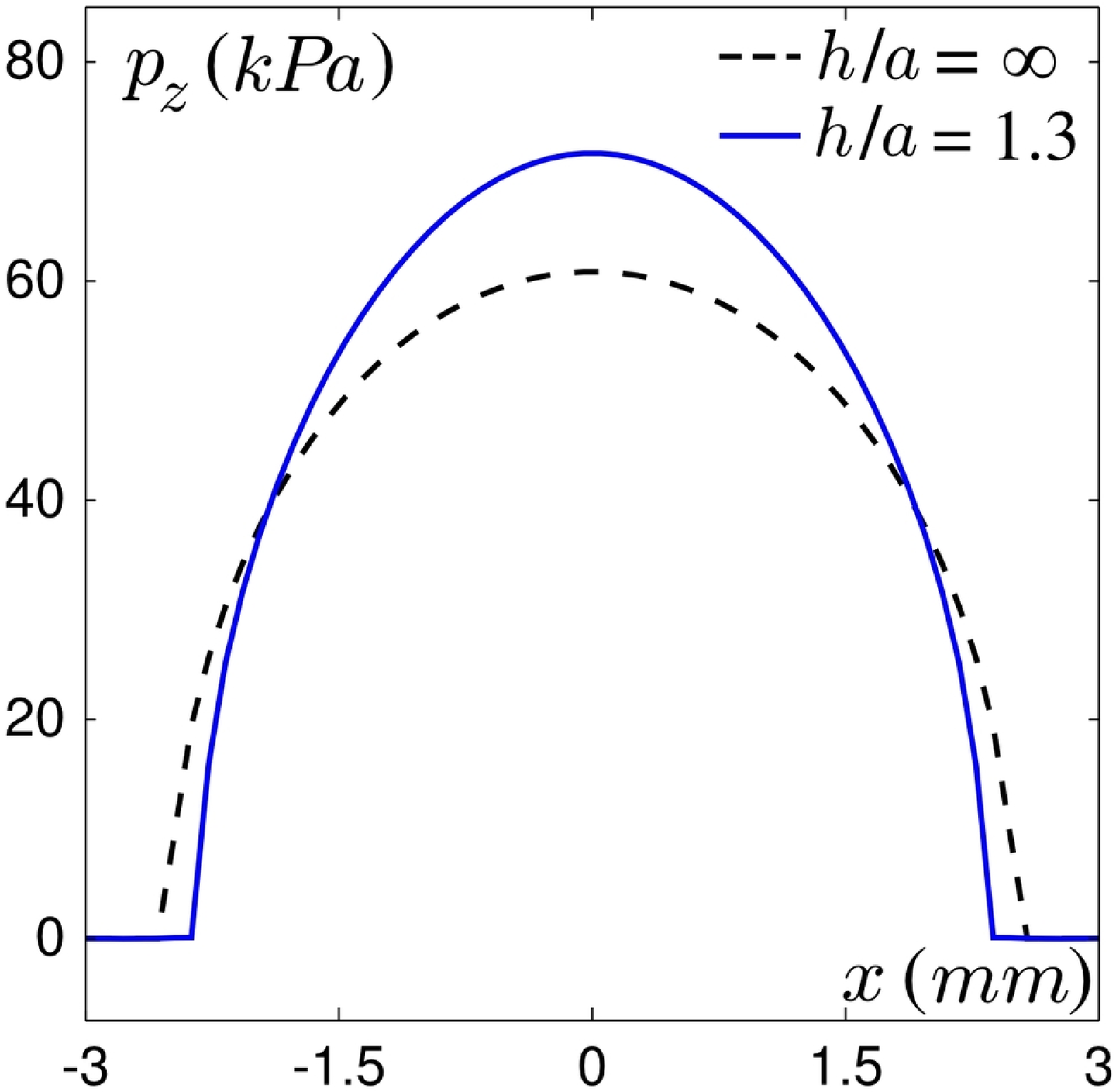}
\includegraphics[height=0.48\columnwidth]{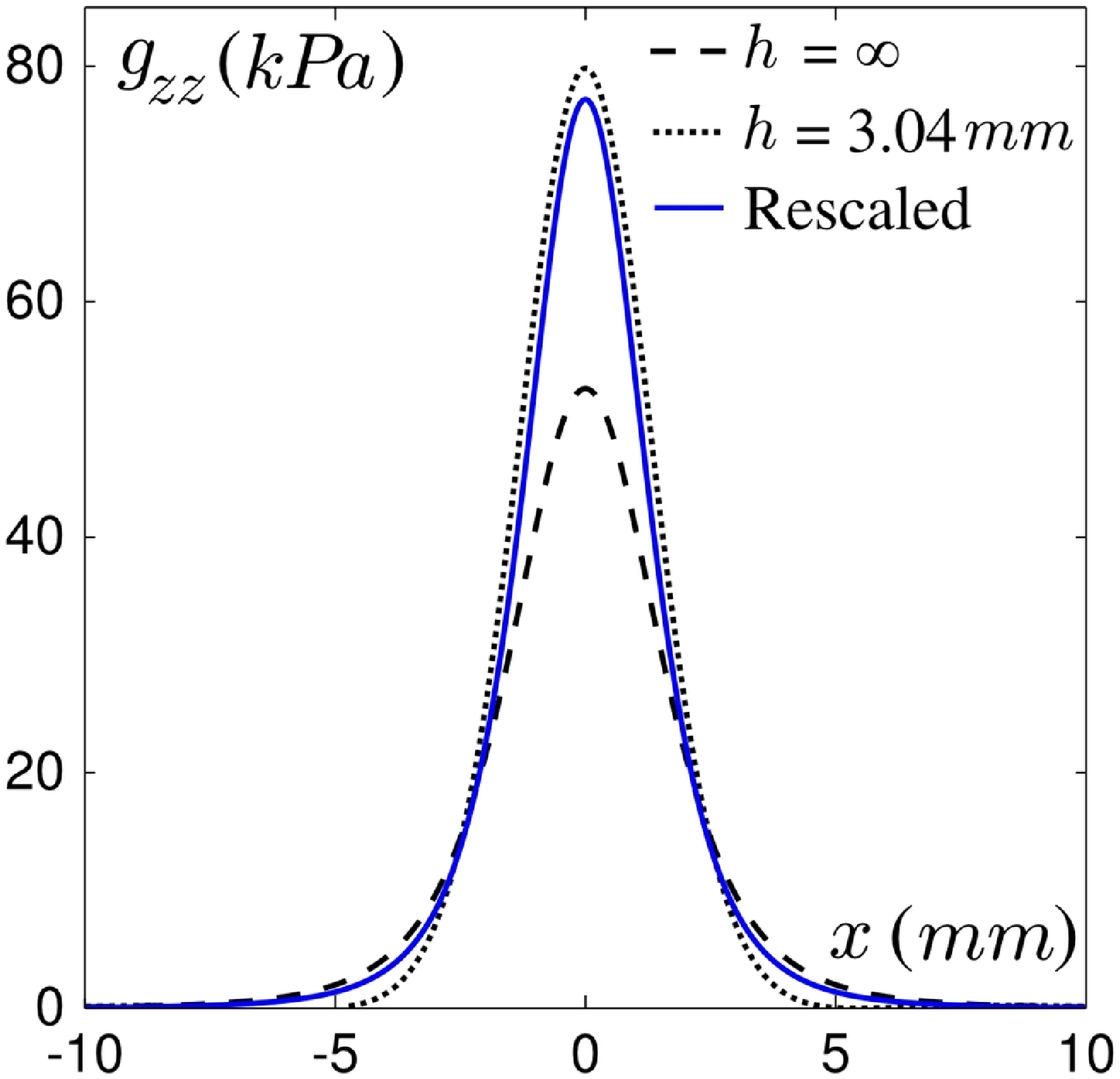}
\caption{ Effect of the finite thickness of the elastic layer.
\leg{Left} Sphere-on-plane contact pressure profile calculated for a semi-infinite elastic medium (dashed black line) and an elastic layer of thickness $h$ (blue line) such that $h/a=1.3$ where $a$ is the contact radius. This ratio corresponds to the experimental conditions.
\leg{Right} Normal component $g_{zz}$ of the Green tensor in ($x$, $y=0$, $z=h$): the profile for the semi-infinite layer (dashed line) can be rescaled (blue line) to approximate the profile obtained by finite element simulation for a layer of thickness $h=3.04 mm$ (dotted line).
}
\label{fig:finite_layer}
\end{figure}

\begin{itemize}
\item \textit{(i) Pressure field at the interface.} The interfacial pressure field in a sphere-on-plane contact under normal load is given by Hertz calculation~\cite{johnson1985cm}. This profile however needs to be corrected to take into account the finite thickness of the elastic layer. Fretigny and Chateauminois have recently proposed a numerical solution to this problem~\cite{fretigny2007sfe}. A comparison between the Hertz pressure profiles for an infinite and a finite thickness layer corresponding to our system geometry is shown in figure~\ref{fig:finite_layer}-left. The modified pressure profile has a smaller contact radius and the pressure at the center is higher.

\item \textit{(ii) Local law of friction.} In steady sliding, we postulate a local Coulomb law everywhere in the contact zone, \ie we assume a linear relationship between the tangential stress $p_x$ and the normal stress $p_z$ at the interface:
\be
p_x = \mu_d.p_z,
\ee
\noindent where the dynamical friction coefficient $\mu_d \sim 3$ is the ratio of the global tangential and normal forces $F_x/F_z$. We also assume that this friction-induced tangential stress field does not affect the pressure  field discussed above.

\item \textit{(iii) Stress propagation through the elastic layer.} Under linear elasticity and quasi-static hypothesis, the stress propagation through the elastic layer is entirely given by the Green tensor $\left\{g_{ij}(x,y)\right\}$ which characterizes the stress field at position $(x,y,z=h)$ in response to a unit localized force applied at the surface of the layer at position $(x=0,y=0)$. For a semi-infinite elastic medium, an analytical solution to this tensor exists~\cite{johnson1985cm}. Again, in the present configuration, this result needs to be corrected to take into account the finite thickness of the elastic layer and its attachment to a rigid base. The finite element analysis we have carried out for normal indentation indicates that a reasonable approximation for this correction can be obtained by rescaling the semi-infinite solution by a simple numerical factor $k$, while the spatial coordinate is rescaled by a factor $k^{-1/2}$ to ensure force conservation. This is shown in fig.~\ref{fig:finite_layer}-right which compares the result of the finite-element calculation with the rescaled semi-infinite solution with a scale factor $k=1.48$. The same factor yields a correct approximation to the tangential component of the stress (not shown). We make the assumption that the same correction holds for the other components of the Green tensor, \ie for the response to a \textit{tangential} localized force.
\end{itemize}

With these three ingredients, an estimated stress profile along the sensor array axis ($y_i=0$) at depth $z_i=h$ can be derived by convoluting the interfacial stress field with the Green tensor:
\ban
\sigma_z(x_i) =&\iint p_x(x,y) g_{xz}(x-x_i, y, h)dxdy \\
 & + \iint p_z(x,y) g_{zz}(x-x_i, y, h)dxdy, \\
\sigma_x(x_i) =&\iint p_x(x,y) g_{xx}(x-x_i, y, h)dxdy \\
 & + \iint p_z(x,y) g_{zx}(x-x_i, y, h)dxdy,
\ean

\noindent In spite of the numerous assumptions made in this model, fig.~\ref{fig:avg_stress} demonstrates that the predicted stress profiles (black curves) consistently captures the experimental data (colored circles) in both static and dynamic conditions. Note that, for the static case, the observed residual tangential force at rest has been accounted for by postulating a tangential stress field $p_x$ proportional to the normal stress field $p_z$. 

\section{Response to elementary topological defects in dynamic exploration}

\begin{figure}[!t]
\center
\includegraphics[width=0.70\columnwidth]{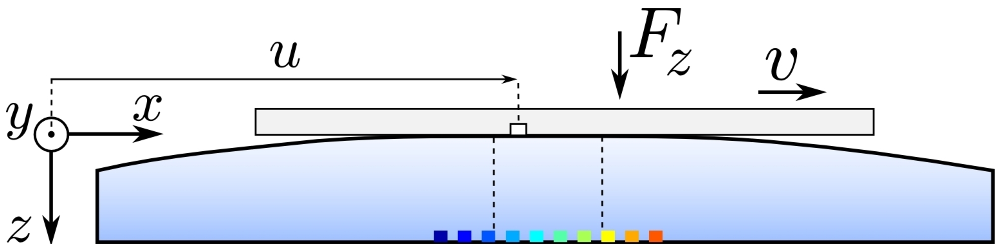}
\includegraphics[height=0.48\columnwidth]{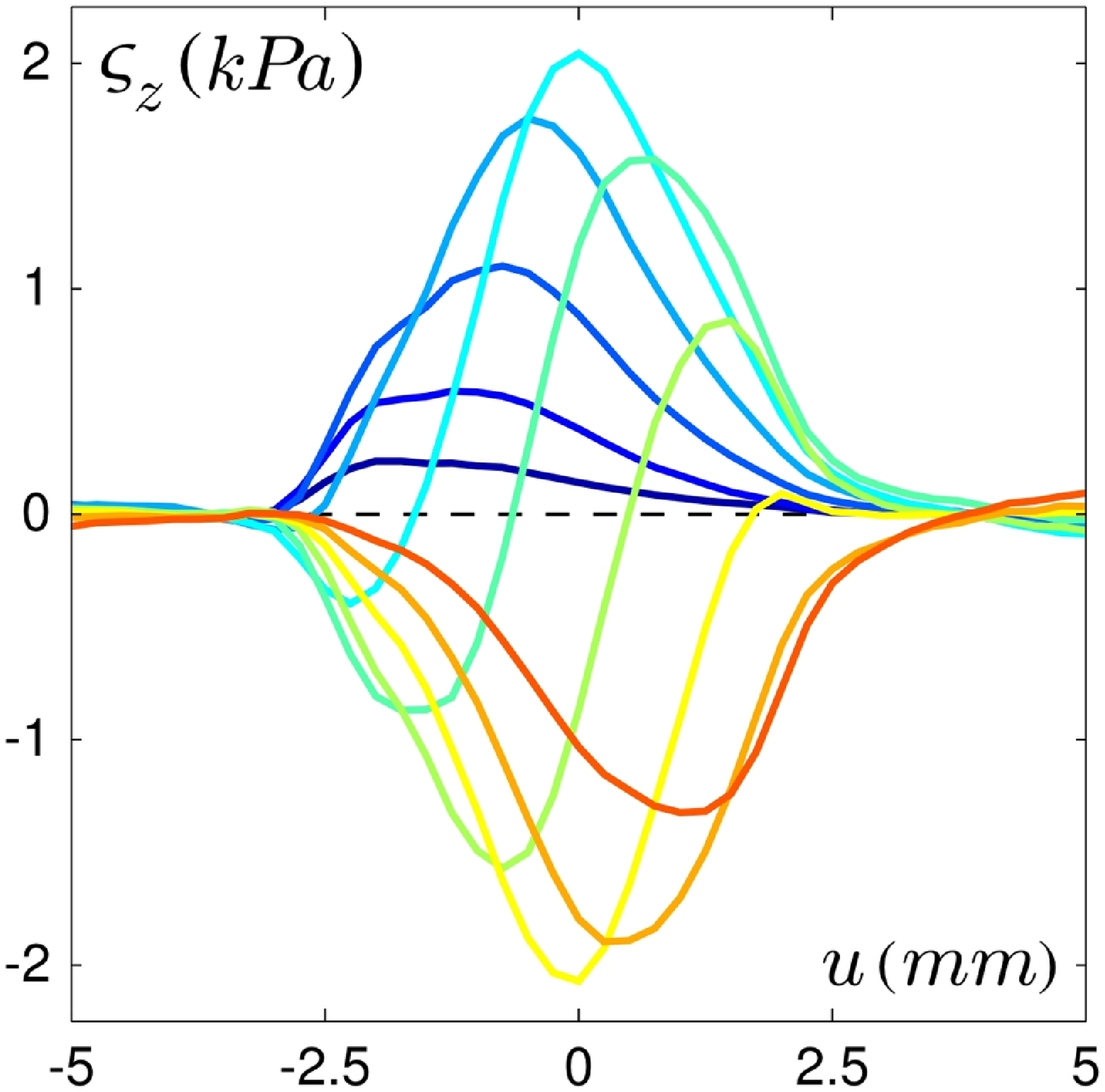}
\includegraphics[height=0.48\columnwidth]{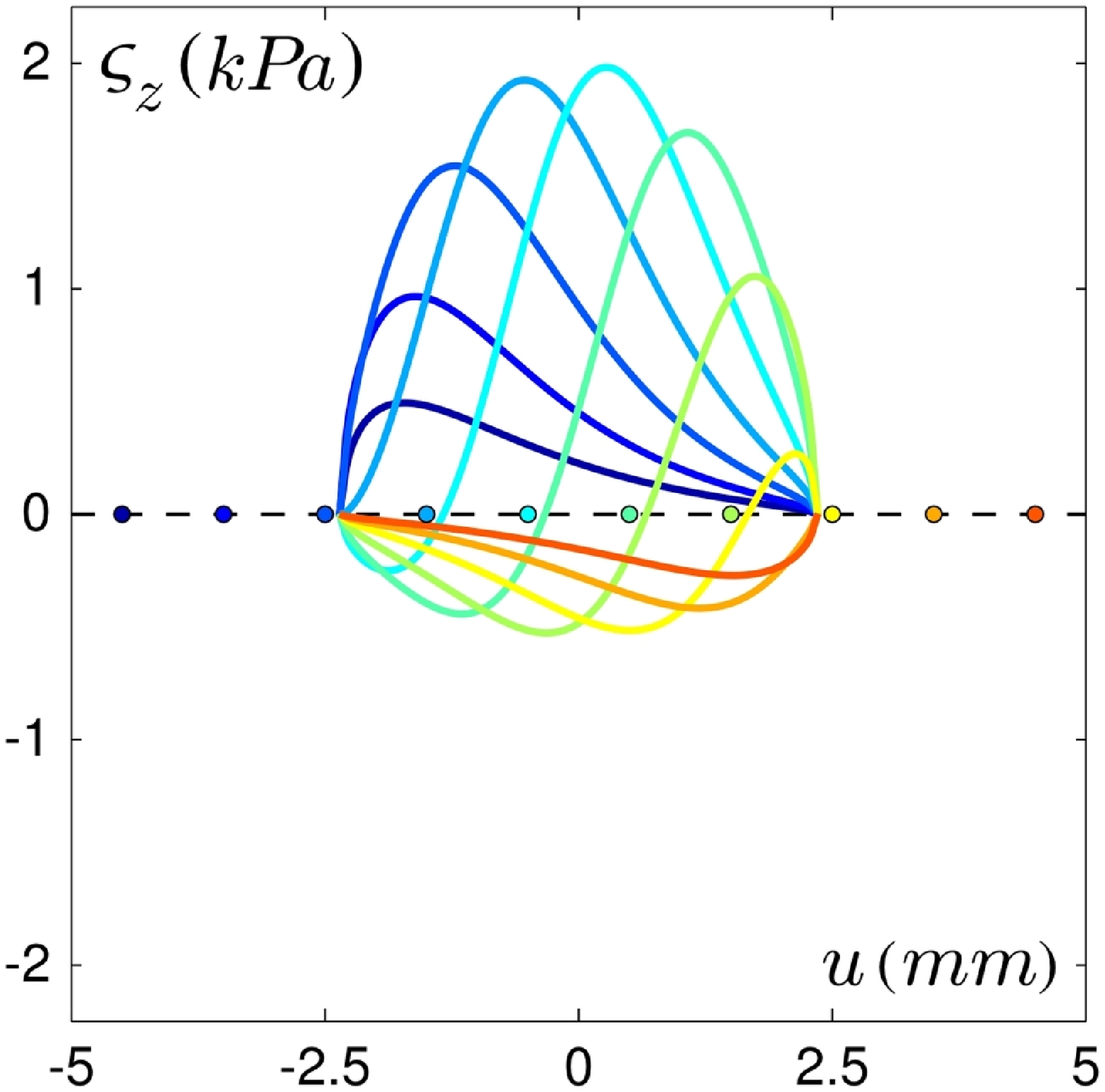}
\includegraphics[height=0.48\columnwidth]{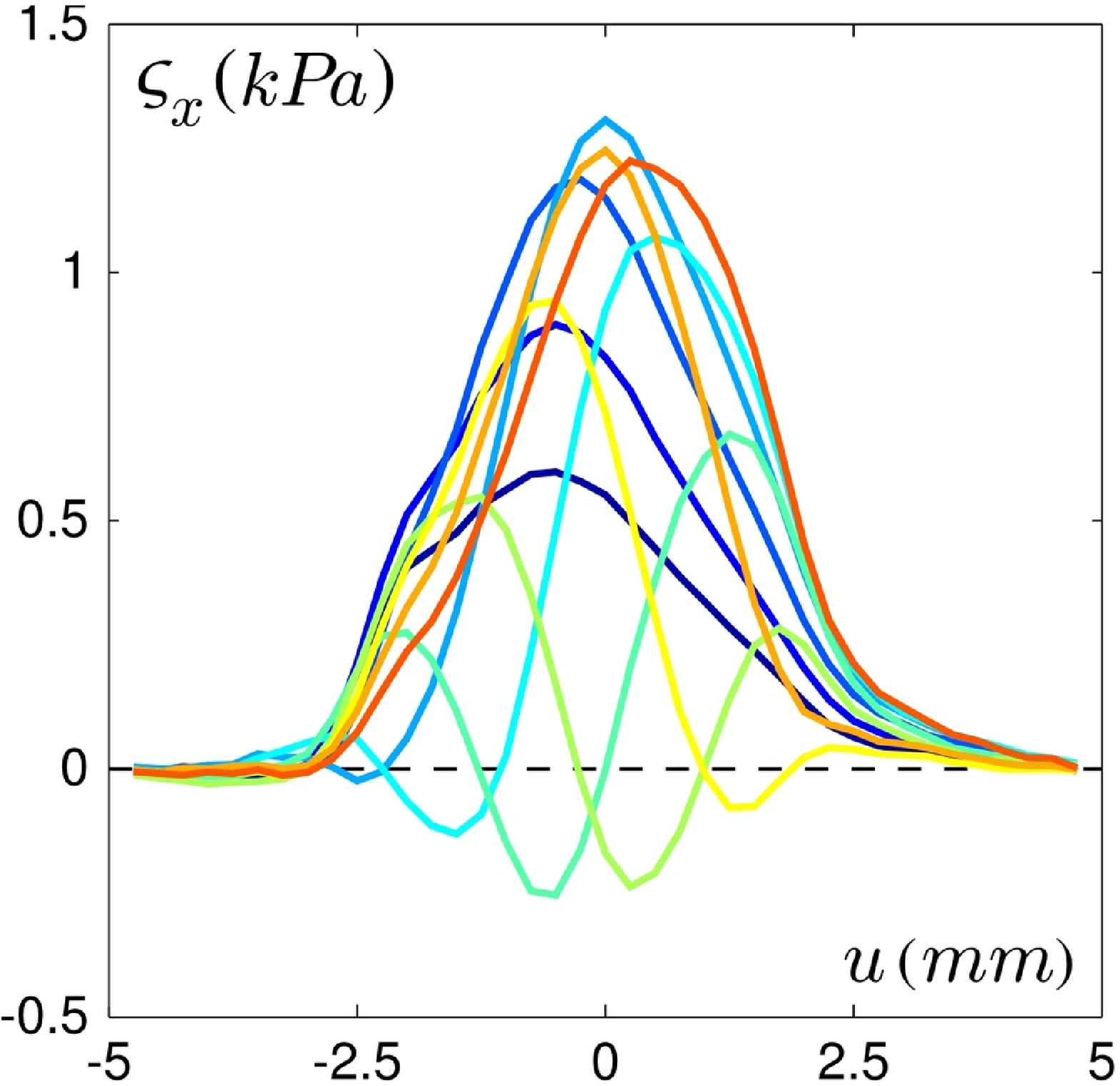}
\includegraphics[height=0.48\columnwidth]{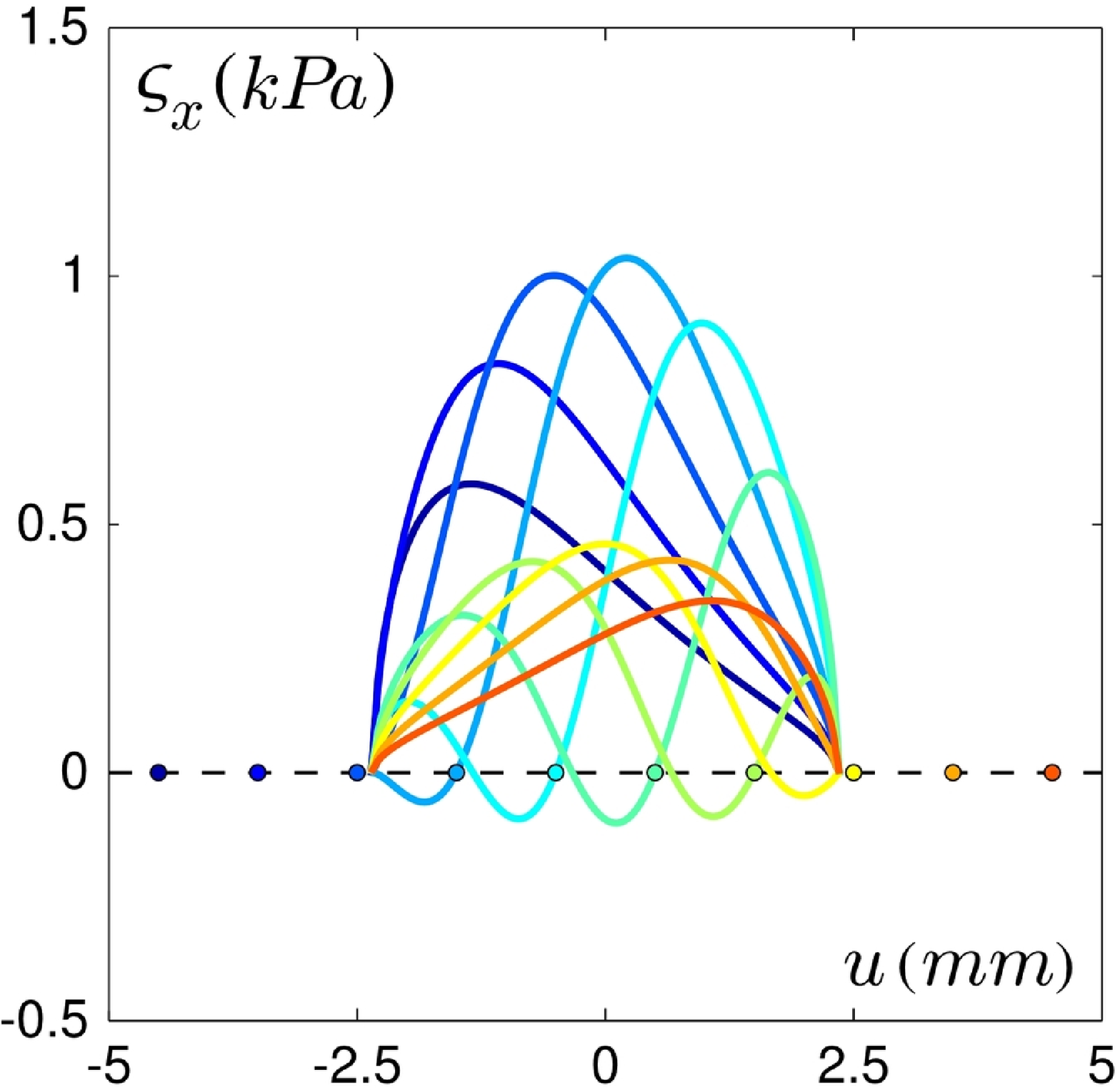}
\caption{ Response to an elementary topological defect scanned across the skin's surface.
\leg{Top}  Sketch of the protocol: a smooth substrate is put into contact with the tactile sensor under a normal load $F_z = 0.8N$. A single defect ($500\mu m$ hole) is scanned along the direction of the sensor's array at constant velocity $v=500\mu m.s^{-1}$.
\leg{Middle} Normal stress fluctuations $\varsigma_z$ measured by the sensors (left) and predicted by the model (right) as functions of the defect's position $u$. The $10$ curves correspond to different probing locations with respect to the contact zone, shown with colored dots on the right plot.
\leg{Bottom} Same as above, but for the tangential component $\varsigma_x$ along the direction of motion.}
\label{fig:fluct}
\end{figure}

We now turn to the response of the biomimetic tactile sensor to the presence of an elementary topological defect on the flat surface. The same Plexiglas plate is patterned with individual holes of diameter $500\mu m$ and depth $500\mu m$ sufficiently well separated in space so that only one hole at most can lay within the contact zone. In static contact, the presence of a hole within the contact zone cannot be detected by the sensor. However, in dynamic exploration, \ie when the hole is scanned across the sensor along the MEMS array axis, the passage of the hole in the contact zone can be easily identified by subtracting the time signal of each sensor with its time-averaged value. We denote $\varsigma_z(u)$ and $\varsigma_x(u)$ the stress variation signals thus defined, with $u$ being the position of the hole along the $x$ direction of motion (see fig.\ref{fig:fluct}-top). Fig.\ref{fig:fluct}-left shows the profiles obtained by averaging over $14$ passages of similar holes, for the ten sensors whose locations range from $-4.5$ to $4.5mm$. Standard deviations are typically $0.6kPa$ for $\varsigma_z$ and $0.4kPa$ for $\varsigma_x$. These dynamic response profiles, unlike the localized force responses shown in fig.~\ref{fig:system_cal}, display very dissimilar shapes depending on the sensor's precise location with respect to the contact zone. The normal stress profiles $\varsigma_z(u)$ evolve from a positive hump on the left-hand side sensors to a negative one on the right-hand side sensors via an intermediate antisymmetric profile (see sensors $\sharp6$ and $7$, in green). In contrast, the shear-stress profiles show a dominant hump on the left and right-hand side sensors, with an intermediate response exhibiting two comparable maxima. Note that the outmost sensors do respond to the defect's passage although they lay outside of the contact zone.

This result can be quantitatively interpreted based on the simple mechanical model presented before. The interfacial pressure field at the location of the hole is strictly zero since the elastic layer does not make contact with the substrate in this area. We further assume that the interfacial stress field in the rest of the contact is unchanged except for an overall uniform rescaling factor to ensure that the integral of the stress over the contact zone remains equal to the imposed overall load $F_z$. Under this assumption, the stress modification induced by a hole located at ($u$,$y=0$) can be approximated, for a sensor located in ($x_i$, $y_i=0$, $z_i=h$), as:
\ban
\varsigma_z(u) \simeq - \left[ p_x(u) g_{xz}(u-x_i) + p_z(u) g_{zz}(u-x_i) \right] S_d\\
\varsigma_x(u) \simeq - \left[ p_x(u) g_{xx}(u-x_i) + p_z(u) g_{zx}(u-x_i) \right] S_d\\
\ean
\noindent where $S_d$ is the hole's surface area. These predictions are shown in the right panels of fig.~\ref{fig:fluct} for ten $1mm$-spaced sensors whose positions are depicted on the horizontal axis with colored circles, allowing for a direct comparison with the experimental measurements. Interestingly, the profiles predicted by the model are not only similar in shape to the measured ones but they are also quantitatively equivalent (compare for instance the profiles for sensor $\sharp5$ in cyan, which can be almost superimposed). The major quantitative difference occurs for the sensors laying on the right-hand side of the contact (from $\sharp6$ to $10$): the measured stress variations amplitudes are up to $4$ times larger than their predicted counterpart, both for the normal and tangential components. This discrepancy may reflect the need to refine our description of the interfacial stress field, in particular to account for the coupling between tangential and normal stresses: the friction-induced tangential stress field produces a torque on the elastic layer which is balanced by a dissymmetrization of the interfacial pressure profile\cite{scheibert2009sfa}. This effect, ignored in our simple model, is expected to increase the signal experienced by the sensors sitting at the front of the contact.

\section{Conclusion and perspectives}


This study constitutes the first step towards the comprehension of mechanical transduction of roughness information in realistic exploratory conditions. It allowed us to exhibit individual sensors' response to the scanning of elementary topological defects. These dynamic response functions should enable one to predict the linear response to any number of such holes patterned on a smooth substrate. By comparing such predictions to actual measurements, one may in turn characterize possible corrections that would reflect the non-linear properties of solid frictional interfaces.

A central issue in artificial touch perception is the implementation of an inversion method, \ie a signal processing procedure allowing one to consistently extract a representation of the probed environment from the measured sensors' responses. A naive strategy would consist in considering each sensor independently and extract from their signal, based on their intrinsic characteristics, the physical information of the substrate that lay within their sensitive area. Our results show that such an approach is likely to fail: in realistic conditions of exploration, the contact zone between the surface and the sensor cannot be known with great precision but the response of each sensor will critically depends on their position with respect to the contact zone. This difficulty is reminiscent of the problem of face recognition in image processing for instance, where variability in the lighting conditions needs to be accounted for in the signal analysis. This analogy suggests that in tactile exploration as well, the signal processing should include a pre-treatment stage based on multi-sensors signal analysis to account for contact stress heterogeneities at the skin/substrate interface. 



\section*{Acknowledgment}
The authors would like to thank A. Chateauminois for the computation of the modified Hertz stress field and J. Scheibert and J. Frelat for their help on the finite element simulation.

\bibliographystyle{IEEEtran}
\bibliography{biblio_finger}

\begin{IEEEbiography}[{\includegraphics[width=1in, height=1.25in, clip, keepaspectratio]{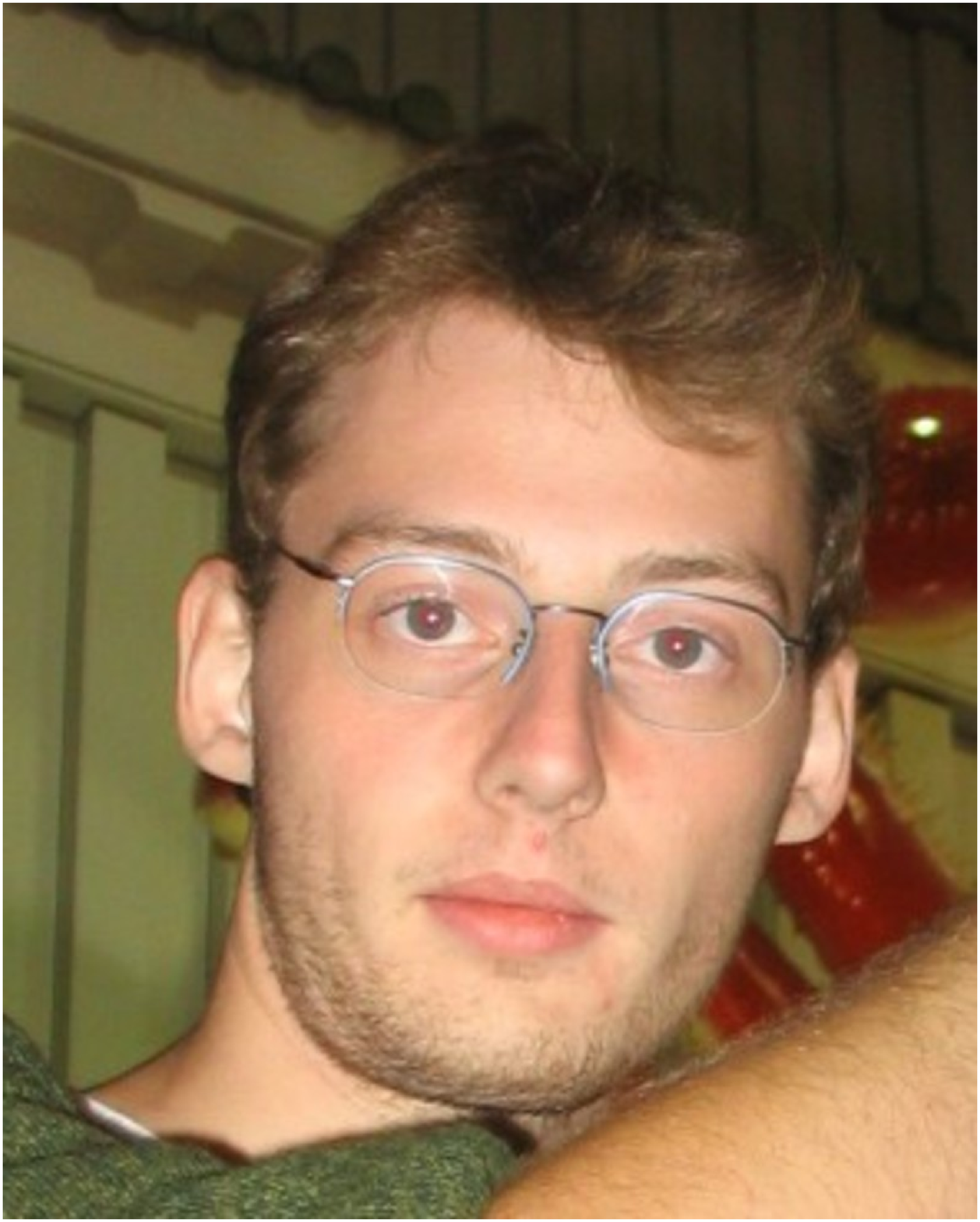}}]{Dr. Raphaël Candelier}
is a post-doc fellow in the sys-bio team of the LPS-ENS, France. He graduated the ENS/EHESS master of Cognitive Sciences, studied psychoacoustics at the IML (Doshisha U., Kyoto) and psycholinguistics at the LSCP (ENS, Paris). He was a member of the Starflag project, an European experimental work aimed at understanding the rules of collective behavior. Nevertheless his major is Physics, which he studied at the ESPCI and during his PhD in the GIT-SPEC (CEA, Saclay), where he specialized in statistical physics.
\end{IEEEbiography}

\begin{IEEEbiography}[{\includegraphics[width=1in, height=1.25in, clip, keepaspectratio]{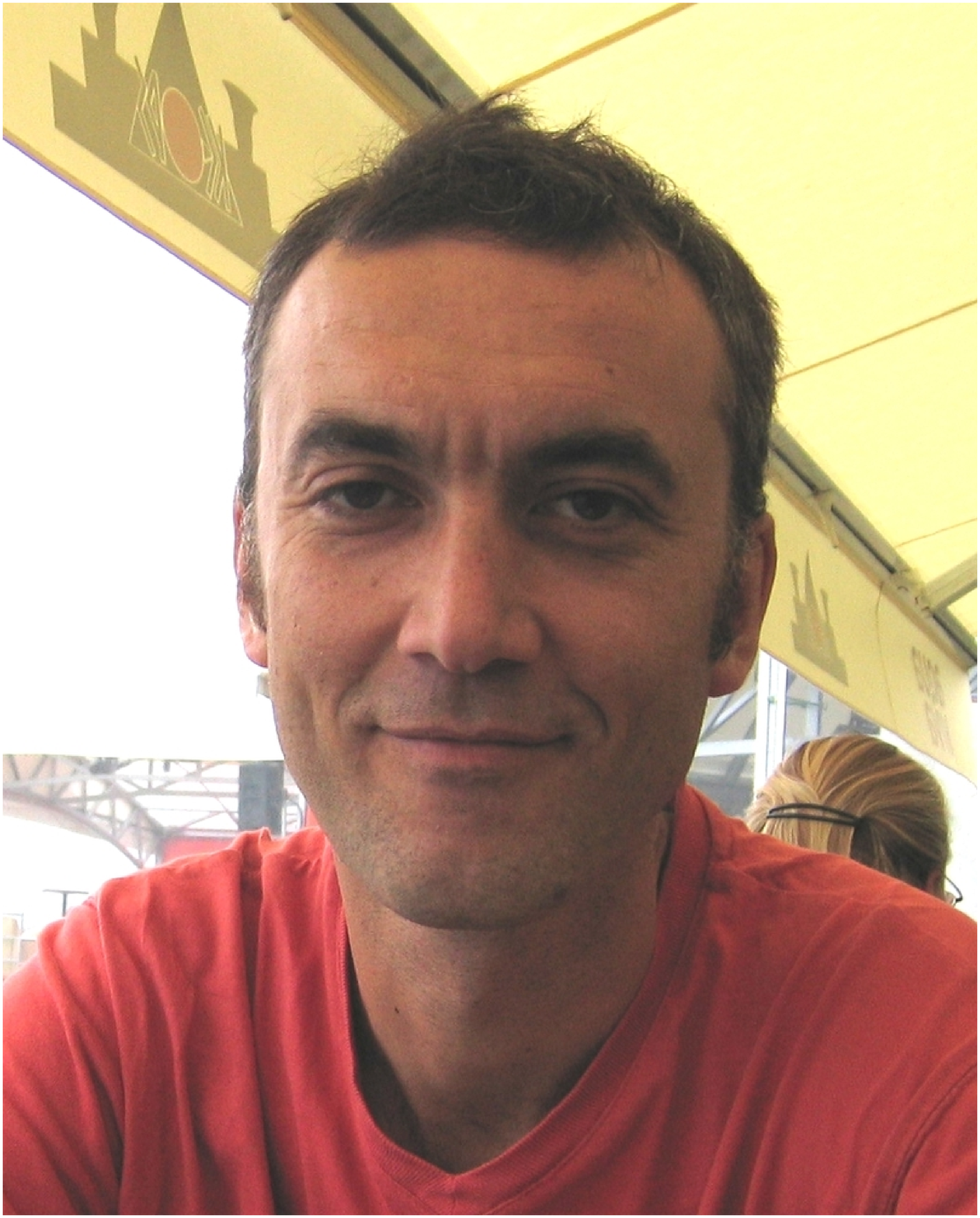}}]{Dr. Georges Debrégeas}
is a CNRS experimental physicist. He received his PhD in 1997, from UPMC. He then joined the James Frank Institute in Chicago as a Post-Doc. In 1999, he got a position as a CNRS researcher, first in Strasbourg (ICS) then in College de France in Paris (LFO), and finally in ENS in 2004. In the earlier years of his career, he studied the rheological properties of various complex systems, such as granular materials and liquid foams. More recently, he got interested in the mechanics of contact and friction and its application to tactile perception in both humans and rodents.
\end{IEEEbiography}

\begin{IEEEbiography}[{\includegraphics[width=1in, height=1.25in, clip, keepaspectratio]{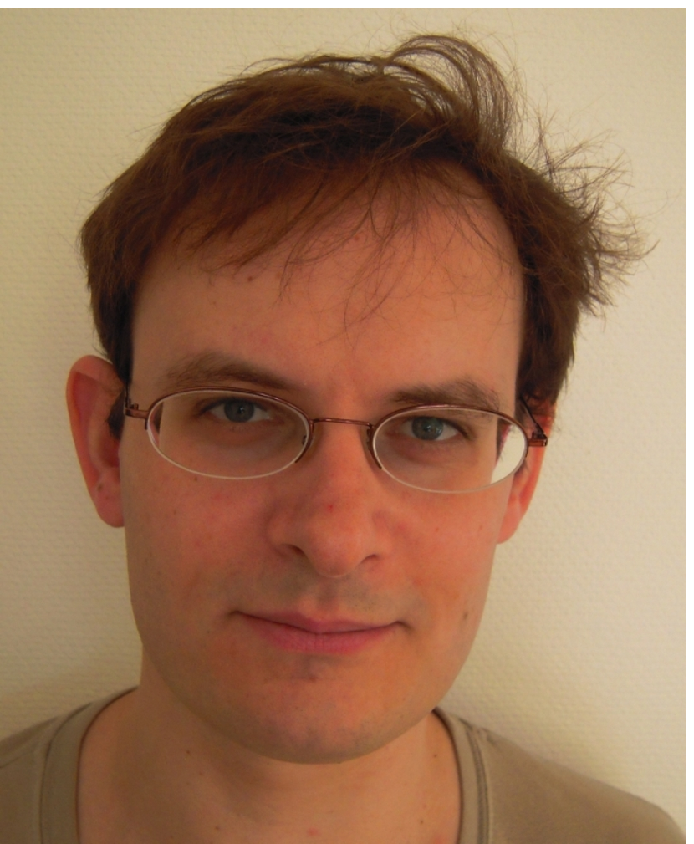}}]{Dr. Alexis Prevost}
is a CNRS experimental physicist whose areas of expertise range from wetting phenomena and granular materials to friction physics. He received his PhD degree in 1999 from the Univ. of Paris XI, a work completed at the LPS-ENS in Paris, France. In 2002, he joined the Physics Department of Georgetown Univ. as a postdoctoral fellow to study granular media dynamics. Since 2002, he has been holding a full-time CNRS researcher position, at LFO-College de France in Paris until 2005, and at LPS-ENS ever since. He now studies the contact mechanics of elastomers on rigid surfaces. He has been developing contact imaging techniques and direct stress measurements using micro-force sensors, in particular to gain a better understanding of the mechanical transduction of the tactile information involved in the human tactile perception.
\end{IEEEbiography}

\end{document}